\documentclass[prd,preprintnumbers]{revtex4}

\usepackage{amsmath}
\usepackage{amsfonts}
\usepackage{graphicx}

\begin{document}
\tolerance=5000
\def\pp{{\, \mid \hskip -1.5mm =}}
\def\cL{{\cal L}}
\def\be{\begin{equation}}
\def\ee{\end{equation}}
\def\bea{\begin{eqnarray}}
\def\eea{\end{eqnarray}}
\def\tr{{\rm tr}\, }
\def\nn{\nonumber \\}
\def\e{{\rm e}}

\preprint{YITP-05-14}

\title{Gauss-Bonnet dark energy}

\author{Shin'ichi Nojiri}\email{snojiri@yukawa.kyoto-u.ac.jp, 
nojiri@cc.nda.ac.jp}
\affiliation{Department of Applied Physics,
National Defence Academy,
Hashirimizu Yokosuka 239-8686, Japan}
\author{Sergei D.~Odintsov\footnote{also at 
Lab. for Fundamental Studies,
Tomsk State Pedagogical University,
634041 Tomsk, RUSSIA}}\email{odintsov@ieec.fcr.es}
\affiliation{Instituci\`o Catalana de Recerca i Estudis
Avan\c{c}ats (ICREA)  and Institut d'Estudis Espacials de Catalunya
(IEEC/ICE),
Edifici Nexus, Gran Capit\`a 2-4, 08034 Barcelona, Spain}
\author{Misao Sasaki}\email{misao@yukawa.kyoto-u.ac.jp}
\affiliation{Yukawa Institute for Theoretical Physics,
Kyoto University, Kyoto 606-8502, Japan}


\begin{abstract}

We propose the Gauss-Bonnet dark energy model inspired by string/M-theory
where standard gravity with scalar contains additional scalar-dependent
coupling with Gauss-Bonnet invariant. It is demonstrated that effective
phantom (or quintessence) phase of late universe may occur in the
presence of such term  when the scalar is phantom or for non-zero
  potential (for canonical scalar).
However, with the increase of the curvature the GB term may become dominant so 
that phantom phase is transient and $w=-1$ barrier
may be passed. Hence, the current acceleration of the universe
may be caused by mixture of scalar phantom and (or) potential/stringy effects. 
It is
remarkable that scalar-Gauss-Bonnet coupling acts against
the Big Rip occurence  in phantom cosmology.

\end{abstract}


\maketitle

\section{Introduction}

It became clear recently that late-time dynamics of the current
accelerated universe is governed by the mysterious dark energy.
The interpretation of the astrophysical observations indicates
that such dark energy fluid (if it is fluid!) is characterized by
the negative pressure and its equation of state parameter $w$ lies very
close to $-1$ (most probably below of it). Quite possible that it may be
oscillating around $-1$. It is extremely difficult to present
the completely satisfactory theory of the dark energy (also due
to lack of all requiered astrophysical data), especially in
the case of (oscillating) $w$ less than $-1$. (For instance,
thermodynamics is quite strange there with possible negative entropy
\cite{brevik}).

The successful dark energy theory may be searched in string/M-theory.
Indeed, it is quite possible that some unusual gravity-matter
couplings predicted by the fundamental theory may become important at
current, low-curvature universe (being not essential in intermediate epoch
from strong to low curvature). For instance, in the study of
string-induced gravity near to initial singularity the role of Gauss-Bonnet 
(GB)
coupling with scalar was quite important for
ocurrence of non-singular cosmology \cite{ART,nick} (for account
of dilaton and higher order corrections near to initial singularity,
see also \cite{modulus}). The present paper is devoted to the study
of the role of GB coupling with the scalar field to the late-time
universe. It is explicitly demonstrated that such term itself can not induce
the effective phantom late-time universe if the scalar is canonical in the 
absence of potential term. It may produce the
effective quintessence (or phantom) era, explaining the current acceleration
only when the scalar is phantom or when the scalar is canonical with non-zero 
potential. It is interesting that it may
also have the important
impact to the Big Rip singularity \cite{bigrip}, similarly to quantum effects
\cite{nojiri,tsujikawa}, preventing it in the standard phantom cosmology. Note 
that we concentrate
mainly on the exponential scalar-GB coupling and exponential scalar
potential, while the consideration of other types of such functions and
their role in late time cosmology will be considered elsewhere.

\section{The accelerated universe from scalar-GB gravity}

We consider a model of the scalar field $\phi$ coupled with gravity.
As a stringy correction, the term proportional to the GB
invariant $G$ is added:
\be
\label{GB1}
G=R^2 - 4 R_{\mu\nu} R^{\mu\nu} + R_{\mu\nu\rho\sigma} R^{\mu\nu\rho\sigma}\ .
\ee
The starting action is given by
\be
\label{GB2}
S=\int d^4x \sqrt{-g}\left\{ \frac{1}{2\kappa^2}R
    - \frac{\gamma}{2}\partial_\mu \phi \partial^\mu \phi
    - V(\phi) + f(\phi) G\right\}\ .
\ee
Here $\gamma=\pm 1$. For the canonical scalar, $\gamma=1$ but at least
when GB term is not included, the scalar behaves as phantom only
when $\gamma=-1$ \cite{caldwell} showing in this case the properties
similar to quantum field \cite{no}. In analogy with model \cite{coupled}
where also non-trivial coupling of scalar Lagrangian with some power of
curvature was considered, one may expect that such GB coupling term may be
relevant for the explanation of dark energy dominance.

By the variation over $\phi$, we obtain
\be
\label{GB3}
0=\gamma \nabla^2 \phi - V'(\phi) + f'(\phi) G\ .
\ee
On the other hand, the variation over the metric $g_{\mu\nu}$ gives
\bea
\label{GB4}
0&=& \frac{1}{\kappa^2}\left(- R^{\mu\nu} + \frac{1}{2} g^{\mu\nu} R\right)
      + \gamma \left(\frac{1}{2}\partial^\mu \phi \partial^\nu \phi
      - \frac{1}{4}g^{\mu\nu} \partial_\rho \phi \partial^\rho \phi \right)
   + \frac{1}{2}g^{\mu\nu}\left( - V(\phi) + f(\phi) G \right) \nn
&&    -2 f(\phi) R R^{\mu\nu} + 2 \nabla^\mu \nabla^\nu \left(f(\phi)R\right)
      - 2 g^{\mu\nu}\nabla^2\left(f(\phi)R\right) \nn
&& + 8f(\phi)R^\mu_{\ \rho} R^{\nu\rho}
      - 4 \nabla_\rho \nabla^\mu \left(f(\phi)R^{\nu\rho}\right)
      - 4 \nabla_\rho \nabla^\nu \left(f(\phi)R^{\mu\rho}\right) \nn
&& + 4 \nabla^2 \left( f(\phi) R^{\mu\nu}  \right)
+ 4g^{\mu\nu} \nabla_{\rho} \nabla_\sigma \left(f(\phi) R^{\rho\sigma} \right)
   - 2 f(\phi) R^{\mu\rho\sigma\tau}R^\nu_{\ \rho\sigma\tau}
+ 4 \nabla_\rho \nabla_\sigma \left(f(\phi) R^{\mu\rho\sigma\nu}\right) \ .
\eea
By using the identities obtained from the Bianchi identity
\bea
\label{GB5}
\nabla^\rho R_{\rho\tau\mu\nu}&=& \nabla_\mu R_{\nu\tau} - \nabla_\nu
R_{\mu\tau}\ ,\nn
\nabla^\rho R_{\rho\mu} &=& \frac{1}{2} \nabla_\mu R\ , \nn
\nabla_\rho \nabla_\sigma R^{\mu\rho\nu\sigma} &=&
\nabla^2 R^{\mu\nu} - {1 \over 2}\nabla^\mu \nabla^\nu R
+ R^{\mu\rho\nu\sigma} R_{\rho\sigma}
      - R^\mu_{\ \rho} R^{\nu\rho} \ ,\nn
\nabla_\rho \nabla^\mu R^{\rho\nu}
+ \nabla_\rho \nabla^\nu R^{\rho\mu}
&=& {1 \over 2} \left(\nabla^\mu \nabla^\nu R
+ \nabla^\nu \nabla^\mu R\right)
      - 2 R^{\mu\rho\nu\sigma} R_{\rho\sigma}
+ 2 R^\mu_{\ \rho} R^{\nu\rho} \ ,\nn
\nabla_\rho \nabla_\sigma R^{\rho\sigma} &=& {1 \over 2} \Box R \ ,
\eea
one can rewrite (\ref{GB4}) as
\bea
\label{GB4b}
0&=& \frac{1}{\kappa^2}\left(- R^{\mu\nu} + \frac{1}{2} g^{\mu\nu} R\right)
      + \gamma \left(\frac{1}{2}\partial^\mu \phi \partial^\nu \phi
      - \frac{1}{4}g^{\mu\nu} \partial_\rho \phi \partial^\rho \phi \right)
   + \frac{1}{2}g^{\mu\nu}\left( - V(\phi) + f(\phi) G \right) \nn
&&    -2 f(\phi) R R^{\mu\nu} + 4f(\phi)R^\mu_{\ \rho} R^{\nu\rho}
   -2 f(\phi) R^{\mu\rho\sigma\tau}R^\nu_{\ \rho\sigma\tau}
+4 f(\phi) R^{\mu\rho\sigma\nu}R_{\rho\sigma} \nn
&& + 2 \left( \nabla^\mu \nabla^\nu f(\phi)\right)R
      - 2 g^{\mu\nu} \left( \nabla^2f(\phi)\right)R
   - 4 \left( \nabla_\rho \nabla^\mu f(\phi)\right)R^{\nu\rho}
      - 4 \left( \nabla_\rho \nabla^\nu f(\phi)\right)R^{\mu\rho} \nn
&& + 4 \left( \nabla^2 f(\phi) \right)R^{\mu\nu}
+ 4g^{\mu\nu} \left( \nabla_{\rho} \nabla_\sigma f(\phi) \right) R^{\rho\sigma}
- 4 \left(\nabla_\rho \nabla_\sigma f(\phi) \right) R^{\mu\rho\nu\sigma} \ .
\eea
The above expression is valid in arbitrary spacetime dimensions.
In four dimensions, the terms proportional to $f(\phi)$ without derivatives, 
are
cancelled with each other and vanish since the GB invariant is a
total derivative in four dimensions.

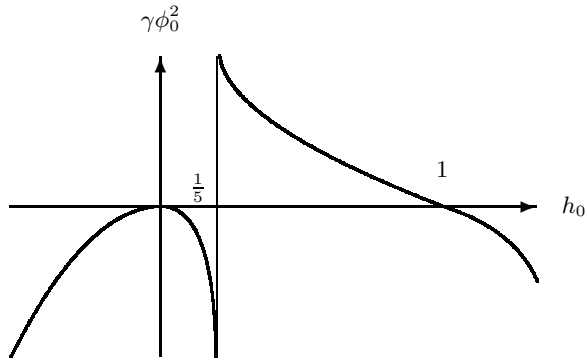
\begin{figure}
\begin{center}
\unitlength=0.5mm
\begin{picture}(150,100)
\thicklines
\put(10,40){\vector(1,0){140}}
\put(160,40){\makebox(0,0){$h_0$}}
\put(50,0){\vector(0,1){80}}
\put(50,90){\makebox(0,0){$\gamma\phi_0^2$}}

\qbezier(10,0)(30,40)(50,40)
\qbezier(50,40)(64,40)(64.5,0)
\qbezier(65.5,80)(68,62)(125,40)
\qbezier(125,40)(144,34)(150,20)
\put(60,45){\makebox(0,0){$\frac{1}{5}$}}
\put(125,50){\makebox(0,0){$1$}}

\thinlines
\put(65,0){\line(0,1){80}}

\end{picture}
\end{center}
\caption{\label{Fig1} The qualitative behavior of $\phi_0^2$ versus $h_0$ from
(\ref{GB18}).}
\end{figure}

The starting Friedmann-Robertson-Walker (FRW) universe metric is:
\be
\label{GB5b}
ds^2=-dt^2 + a(t)^2 \sum_{i=1}^3  \left(dx^i\right)^2\ ,
\ee
where
\bea
\label{GB6}
&& \Gamma^t_{ij}=a^2 H \delta_{ij}\ ,\quad
\Gamma^i_{jt}=\Gamma^i_{tj}=H\delta^i_{\ j}\ ,
\quad R_{itjt}=-\left(\dot H + H^2\right)\delta_{ij}\ ,\quad
R_{ijkl}=a^4 H^2\left(\delta_{ik} \delta_{lj} - \delta_{il} \delta_{kj}\right)\ 
,\nn
&& R_{tt}=-3\left(\dot H + H^2\right)\ ,\quad R_{ij}= a^2 \left(\dot H
+ 3H^2\right)\delta_{ij}\ ,\quad R= 6\dot H + 12 H^2\ , \quad
\mbox{other components}=0\ ,
\eea
(here the Hubble rate $H$ is defined by $H=\dot a/ a$).
Assuming $\phi$ only depends on time, the $(\mu,\nu)=(t,t)$-component in 
(\ref{GB4})
has the following simple form:
\be
\label{GB7}
0=-\frac{3}{\kappa^2}H^2 + \frac{\gamma}{2}{\dot \phi}^2 + V(\phi) - 24 \dot 
\phi f'(\phi) H^3 \ .
\ee
On the other hand, Eq.(\ref{GB2}) becomes:
\be
\label{GB8}
0=-\gamma\left(\ddot\phi + 3H\dot\phi\right) - V'(\phi) + 24 f'(\phi)
\left(\dot H H^2 + H^4\right)\ .
\ee
We now consider the case that $V(\phi)$ and $f(\phi)$ are given as
exponents with the constant parameters $V_0$, $f_0$, and $\phi_0$
\be
\label{GB9}
V=V_0\e^{-\frac{2\phi}{\phi_0}}\ ,\quad f(\phi)=f_0 \e^{\frac{2\phi}{\phi_0}}\ 
.
\ee
Assume that the scale factor behaves as $a=a_0t^{h_0}$ ( power law).
In case that $h_0$ is negative, this scale factor does not correspond to 
expanding
universe but it corresponds to shrinking one.
If one changes the direction of time as $t\to -t$, the expanding universe whose
scale factor is given by $a=a_0(-t)^{h_0}$ emerges. In this expression, 
however, since
$h_0$ is not always an integer, $t$ should
be negative so that the scale factor should be real. To avoid the apparent
difficulty, we may further shift
the origin of the time as $t\to -t \to t_s - t$. Then the time $t$ can be
positive as long as $t<t_s$.
Hence, we can propose
\be
\label{GB10}
H=\frac{h_0}{t}\ ,\quad \phi=\phi_0 \ln \frac{t}{t_1}\ ,
\ee
when $h_0>0$ or
\be
\label{GB11}
H=-\frac{h_0}{t_s - t}\ ,\quad \phi=\phi_0 \ln \frac{t_s - t}{t_1}\ ,
\ee
when $h_0<0$, with an undetermined constant $t_1$. By the assumption
(\ref{GB10}) or (\ref{GB11}), one obtains
\be
\label{GB12}
0=-\frac{3h_0^2}{\kappa^2} + \frac{\gamma\phi_0^2}{2} + V_0 t_1^2 - \frac{48
f_0 h_0^3}{t_1^2}\ ,
\ee
from (\ref{GB7}) and
\be
\label{GB13}
0=\gamma\left( 1 - 3h_0 \right)\phi_0^2 + 2V_0 t_1^2
  + \frac{48 f_0 h_0^3}{t_1^2}\left(h_0 - 1\right)\ ,
\ee
from (\ref{GB8}). Using (\ref{GB12}) and (\ref{GB13}), it follows
\bea
\label{GB15}
V_0 t_1^2&=& - \frac{1}{\kappa^2\left(1 + h_0\right)}\left\{3h_0^2 \left( 1 - 
h_0\right)
+ \frac{ \gamma \phi_0^2 \kappa^2 \left( 1 - 5 h_0\right)}{2}\right\}\ ,\nn
\frac{48 f_0 h_0^2}{t_1^2}&=& - \frac{6}{\kappa^2\left( 1 + 
h_0\right)}\left(h_0
    - \frac{\gamma \phi_0^2 \kappa^2}{2}\right)\ .
\eea
The second Eq.(\ref{GB15}) shows that if $-1<h_0<0$ and $\gamma=1$, $f_0$
should be negative.
Without the GB term, that is, $f_0=0$, a well known result follows:
\be
\label{GB16o}
h_0=\frac{\gamma \phi_0^2 \kappa^2}{2}\ .
\ee
Since the equation of state parameter $w$ is given by
\be
\label{GB16}
w=-1 + \frac{2}{3h_0}\ ,
\ee
if $h_0<0$ ($h_0>0$), $w<-1$ ($w>-1$). Eqs.(\ref{GB15}) indicate that even if
$\gamma=1$, with the proper choice of
parameters  $h_0$ can be negative or $w<-1$. Even if $\gamma>0$, when $h_0<-1$,
$V_0$ is positive, which means that the potential $V(\phi)$ is bounded below.
As a special case we consider
\be
\label{GB18}
\phi_0^2 = - \frac{6h_0^2\left( 1- h_0\right)}{\gamma \left(1 - 5 
h_0\right)\kappa^2}\ ,
\ee
which gives $V(\phi)=0$. In order that $\phi_0$ could be real, one has
\be
\label{GB19}
\frac{1}{5}<h_0<1\ ,\quad \mbox{when}\ \gamma=1\ , \quad \mbox{or} \quad
h_0>\frac{1}{5}\ \mbox{or}\ h_0\geq 1\ .
\ee
In the case (\ref{GB18}), the scalar field $\phi$ is canonical ($\gamma=1$),
and there is no potential $V(\phi)=0$, even if we include the term proportional
to the GB invariant, we cannot obtain the effective phantom cosmological 
solution
with $h_0<0$ or $w<-1$.
Eq.(\ref{GB15}) tells, however, when $\gamma=1$ and $V_0>0$ even if $V_0$ is 
arbitrary small,
if we choose $f_0$ properly, we may obtain the effective phantom.
The qualitative behavior of $\gamma\phi_0^2$ versus $h_0$ when $V_0=0$ is given
in Figure \ref{Fig1}.
There is one positive solution, which may mimic the effective matter with 
$1/5<h_0<1$ when $\gamma=1$.
We also find, when $\gamma=-1$,  there are always three solutions for $h_0$ 
from (\ref{GB18}), one is given by $h_0<0$ and describes the phantom cosmology, 
one is
$h_0>1$ describing the quintessence cosmology, and another corresponds to the 
matter with $0<h_0<1/5$.
Then even if $\gamma=-1$, there appear the solutions describing non-phantom 
cosmology
coresponding the quintessence or matter.

As an example, we consider the case that
\be
\label{GR19}
h_0=-\frac{80}{3}<-1\ ,
\ee
which gives, from (\ref{GB16}),
\be
\label{GR20}
w= - 1.025\ ,
\ee
This is consistent with the observational bounds for effective $w$ (for recent
discussion and complete list of refs., see\cite{lazkoz}).
Then from (\ref{GB15}), one obtains
\be
\label{GR22}
V_0t_1^2 = \frac{1}{\kappa^2}\left( \frac{531200}{231}
+ \frac{403}{154}\gamma \phi_0 \kappa^2 \right)\ ,\quad
\frac{f_0}{t_1^2} = -\frac{1}{\kappa^2}\left( \frac{9}{49280}
+ \frac{27}{7884800}\gamma \phi_0 \kappa^2 \right)\ .
\ee
Therefore even starting from the canonical scalar theory with positive 
potential
before introducing the term proportional to the GB invariant, we may
obtain a solution which reproduces the observed value of $w$ as in
(\ref{GR20}).

In case of the model induced from the string theory \cite{ART}, we have $V_0=0$
($V(\phi)=0$) and
\be
\label{ART1}
\phi_0^2 = \frac{2}{\kappa^2}\ ,
\ee
in (\ref{GB9}). Then Eq. (\ref{GB18}) reduces as
\be
\label{ART2}
3h_0^3 -3 h_0^2 +5 h_0 -1 =0\ ,
\ee
which has only one real solution as
\be
\label{ART3}
h_0=0.223223\ .
\ee
The solution gives
\be
\label{ART4}
w= 1.98654\ .
\ee

There is another  solution of (\ref{GB7}) and (\ref{GB8}) with
(\ref{GB9}). In the solution, $\phi$ and $H$ are constants,
\be
\label{GB28}
\phi=\varphi_0\ ,\quad H=H_0\ ,
\ee
what corresponds to deSitter space.
Using (\ref{GB7}) and (\ref{GB8}) with (\ref{GB9}), one finds
\be
\label{GB29}
H_0^2 = - \frac{\e^{-\frac{2\varphi_0}{\phi_0}}}{8f_0 \kappa^2}\ .
\ee
Therefore in order for  the solution to exist, we may require $f_0<0$. In
(\ref{GB29}), $\varphi_0$ can
be arbitrary. Hence, the Hubble rate $H=H_0$ might be determined
by an initial condition.

In case of the model (\ref{GB9}), the term including the GB invariant
always gives
the contribution in the same order with those from other terms even if the
curvature is small.
This is due to the factor $f(\phi)$, which enhances the contribution when the
curvature is small.

\section{Late-time asymptotic cosmology in scalar-GB gravity and Big Rip
avoidance}

In the following, another model, which is slightly different from (\ref{GB9}),
may be considered:
\be
\label{GB30}
V(\phi)=V_0\e^{-\frac{2\phi}{\phi_0}}\ ,\quad f(\phi)=f_0
\e^{\frac{2\phi}{\alpha\phi_0}}\ ,\quad
\left(\alpha>1\right)\ ,
\ee
Different from the model  (\ref{GB9}), the model (\ref{GB30})
will not be solved exactly.
We can only find the asymptotic qualitative behavior of the solutions.
Nevertheless,
the asymptotic behavior  suggests the existence of the cosmological solution,
where the value of $w$
could vary with time (oscillation) and/or could depend on the curvature.

Assuming the solution behaves as (\ref{GB10}) or (\ref{GB11}), when the
curvature is small, that is
$t$ in (\ref{GB10}) or $t_s - t$ in (\ref{GB11}) is large,
the GB term becomes small and could be neglected since it behaves like
$1/t^{-\frac{2}{\alpha} + 4}$ or $1/\left(t_s - t\right)^{-\frac{2}{\alpha} + 
4}$.
When the curvature is small,
the solution could be given by (\ref{GB16o}), then the effective phantom phase
with $w<-1$ could appear only in case $\gamma=-1<0$.
On the other hand, when the curvature is large, that is
$t$ in (\ref{GB10}) or $t_s - t$ in (\ref{GB11}) is small, the classical
potential could be neglected.
Without the classical potential, by assuming, instead of (\ref{GB10})
\be
\label{GB31}
H=\frac{h_0}{t}\ ,\quad \phi=\alpha \phi_0 \ln \frac{t}{t_1}\ ,
\ee
when $h_0>0$ or
\be
\label{GB32}
H=-\frac{h_0}{t_s - t}\ ,\quad \phi=\alpha \phi_0 \ln \frac{t_s - t}{t_1}\ ,
\ee
when $h_0<0$,  the following equations replace (\ref{GB12}) and (\ref{GB13})
\bea
\label{GB33}
0&=&-\frac{3h_0^2}{\kappa^2} + \frac{\gamma\alpha^2 \phi_0^2}{2} - \frac{48 f_0 
h_0^3}{t_1^2}\ ,\nn
0&=&\gamma\left( 1 - 3h_0 \right)\alpha^2 \phi_0^2  + \frac{48 f_0 
h_0^3}{t_1^2}
\left(h_0 - 1\right)\ .
\eea
By deleting $f_0$ in the above two equations, one gets
\be
\label{GB18B}
\phi_0^2 = -\frac{6h_0^2\left( 1 - h_0\right)}{\gamma \alpha^2 \left( 1 - 5 
h_0\right)\kappa^2}\ ,
\ee
which corresponds to (\ref{GB18}).
Then when $\gamma=1$, the solutions of (\ref{GB18B}) are not qualitatively 
changed from those of
(\ref{GB18}), and there is only one solution $1/5<h_0<1$. On the other hand,
when $\gamma=-1$, since the sign of the r.h.s. in (\ref{GB18}) is changed from 
$\gamma=1$ case,
as clear from FIG.\ref{Fig1}, there are three solutions, corresponding to the 
phantom $h_0<0$ or $w<-1$,
the quintessence $h_0>1$ or $-1<w<-1/3$, and the matter with $0<h_0<1/5$ or 
$w>7/3$.
Then if  the term proportional to the GB invariant in case $\gamma<0$
(which corresponds to a scalar phantom solution without GB term)
is included the effective $w$ can become larger than $-1$ and the Big Rip
singularity might be avoided (see \cite{nojiri} for quantum effects account to 
escape of Big Rip).
That is, in case $\gamma<0$, when the curvature is small as in the current
universe, the GB term is negligible and the potential term dominates, which 
gives the cosmic
acceleration with $w<-1$. Then the curvature increases gradually and the 
universe seems
to tend to the Big Rip singularity \cite{bigrip}. However, when the curvature 
is large,
the GB term  becomes dominant and might prevent the singularity. Hence, in case
$\gamma<0$, the GB term may work against the Big Rip singularity occurence,
like quantum effects \cite{nojiri}.
After the GB term dominates when $\gamma<0$, the curvature turns to become 
smaller.
Then the potential term  dominates again. This might tell that the behavior of 
the universe might
approach to the deSitter sapce with $w=-1$ by the dampted oscillation.
In fact even in the model (\ref{GB30}),
if (\ref{GB28}) is assumed, there is a deSitter solution corresponding to
(\ref{GB28}):
\be
\label{GB29B}
H_0^2 = - \frac{\e^{-\frac{2\varphi_0}{\alpha\phi_0}}}{8f_0 \kappa^2}\ ,
\quad \varphi_0=\frac{\alpha\phi_0}{2\left(1-\alpha\right)}
\ln \left( - \frac{8V_0f_0 \kappa^2}{3}\right)\ .
\ee
In (\ref{GB30}), we have assumed $\alpha>1$. If we consider the case that
$V(\phi)=V_0\e^{-\frac{2\phi}{\phi_0}}$ and 
$f(\phi)=f_0\e^{\frac{2\phi}{\alpha\phi_0}}$
as in (\ref{GB30}) but $0<\alpha<1$, there appears a solution where the term
including GB invariant becomes dominant even if the curvature is small.
By assuming (\ref{GB31}) or (\ref{GB32}),  Eq.(\ref{GB33}) is obtained again.
Hence, when $0<\alpha<1$, the solution where $h_0$ can be positive or $w>-1$ 
even if
$\gamma<0$ (scalar phantom) appears.

On the other hand, if $\gamma>0$ there is no accelerated universe solution with 
$w<-1$.
The parameter $w$ may change with time but $w$ is larger than $-1/3$.
curvature
due to the GB term.
effective equation of
occur eventually.

\section{Discussion}

We  considered essentially two models with exponential couplings given by
(\ref{GB9}) and (\ref{GB30}).
The model (\ref{GB9}) may be considered as the special case corresponding to 
$\alpha=1$.
The main results can be summarized as follows:

\begin{enumerate}
\item $\alpha=1$ case: exactly solvable
\begin{enumerate}
\item $V=0$ case:
When $\gamma=1$, there is only one solution $-1/3<w<7/3$. On the other hand,
when $\gamma=-1$, there are three solutions, corresponding to $w<-1$, 
$-1<w<-1/3$, and $w>7/3$.
\end{enumerate}
\item $\alpha> 1$ case: the potential term dominates for small curvature and 
the GB term for
large one.
\begin{enumerate}
\item $\gamma>0$: The value of $w$ may be time dependent but there is no 
solution
describing acceleration of the universe.
\item $\gamma<0$: There might appear the Big Rip singularity but there might be
a solution asymptotically approaching to the deSitter space.
\end{enumerate}
\item $0<\alpha< 1$ case: the potential term dominates for large curvature and 
the GB term for
small one.
\begin{enumerate}
\item $\gamma>0$: There is no solution describing acceleration of the universe.
\item $\gamma<0$: There  appears the Big Rips singularity.
\end{enumerate}
\end{enumerate}

For the models (\ref{GB9}) and (\ref{GB30}), in case $V_0=0$ (that is, when the 
potential vanishes),
by replacing $\alpha\phi_0$ with $\phi_0$, it follows the two models are 
equivalent.
Especially $\gamma=-1$ case has been well studied and it has been shown that
there are always three effective cosmological phases corresponding to the 
phantom with $h_0<0$
or $w<-1$, the quintessence
with $h_0>1$ or $-1<w<-1/3$, and the matter with $0<h_0<1/5$ or $w>7/3$.
Even if $V_0\neq 0$, the model  (\ref{GB9}) can be solved exactly and the 
solutions
where $h_0$ and therefore $w$ are constants may be found. On the other hand, if
$V_0\neq 0$, in the model (\ref{GB30})  there exist
the solutions where the values of $h_0$ and therefore of $w$ are 
time-dependent.
There could emerge a cosmology,
which behaves as phantom one with $w<-1$ when the curvature is small and as a
usual matter dominated universe with $w>-1$ when the curvature is large.
Moreover, Big Rip singularity does not occur.

Our study indicates that current acceleration may be significally influenced by
stringy/M-theory
effects (terms) which somehow became relevant quite recently (in
cosmological sense). It remains a challenge to construct the consistent
dark energy universe model from
string/M-theory.

\noindent
{\bf Note added:} After the first version of this paper (with some error) 
appeared in hep-th,
the related study, for instance, of the influence of scalar-GB term to Big Rip
has appeared in ref.\cite{CST}.

\section*{Acknowledgments}

This research has been supported in part by JSPS
Grants-in-Aid for Scientific Research, No.~13135208 (S.N.) and
Grants-in-Aid for Scientific Research (S), No.~14102004 (M.S.).

\appendix


\section{Stability of phantom cosmology}

In this appendix, we check the stability of the above solutions.
The following quantities are convenient to introduce:
\be
\label{GB38}
X\equiv \frac{\dot\phi}{H}\ ,\quad Z\equiv H^2 f'(\phi)\ ,\quad
\frac{d}{dN}\equiv a\frac{d}{da}= \frac{1}{H}\frac{d}{dt}\ .
\ee
For simplicity, we also put $\kappa^2$ to be unity.
Then by using (\ref{GB7}) and (\ref{GB8}) with (\ref{GB9}), one finds
\bea
\label{GB39}
\frac{dX}{N}&=&\frac{\gamma^2 \phi_0 X^3 + 2\gamma X\left(8X^2Z - \phi_0\left(3 
+ 52XZ\right)\right)
+ 4\left(\frac{2V_0f_0}{\phi_0Z} + \frac{24V_0f_0X}{\phi_0} + 12Z\left(\phi_0
+ 16\phi_0 XZ - 8X^2Z\right)\right)}{2\phi_0\left(\gamma + 6\gamma XZ + 96 
Z^2\right)}\ ,\\
\label{GB40}
\frac{dZ}{dN}&=&
\frac{Z\left( -\gamma^2\phi_0 X^2 -16Z\left(\frac{2V_0f_0}{\phi_0 Z}
+ 12\left(\phi_0 - X\right)Z\right) + 2\gamma\left(X + 16\phi_0 
XZ\right)\right)}
{\phi_0\left(\gamma + 6\gamma XZ + 96 Z^2\right)}\ .
\eea
For the solution  (\ref{GB10}) or (\ref{GB11}), it follows
\be
\label{GB41}
X=X_0\equiv \frac{\phi_0}{h_0}\ ,\quad Z=Z_0\equiv \frac{2f_0 h_0^2}{\phi_0
t_1^2}\ .
\ee
In terms of $X_0$ and $Z_0$, Eqs.(\ref{GB12}) and (\ref{GB13}) can be rewritten 
as
\bea
\label{GB42}
0&=& - \frac{3}{\kappa^2} + \frac{\gamma X_0^2}{2} + \frac{2V_0 f_0}{\phi_0
Z_0} - 24 Z_0 X_0\ ,\\
\label{GB43}
0&=&\gamma X_0^2 - 3\phi_0 \gamma X_0 + \frac{4V_0f_0}{\phi_0Z_0} + 24 \phi_0
Z_0 - 24 Z_0 X_0\ .
\eea
For the solution (\ref{GB41}), by using (\ref{GB42}) and (\ref{GB43}),
the right hand sides of Eqs.(\ref{GB39}) and (\ref{GB40}) vanish consistently.
We now consider the perturbation around the solution  (\ref{GB41}):
\be
\label{GB44}
X=X_0 + \delta X\ ,\quad Y=Y_0 + \delta Y\ .
\ee
We now only check the stability for $V=0$ ($V_0=0$) case.

Using (\ref{GB39}) and (\ref{GB40}), one obtains
\be
\label{GB45}
\frac{d}{dN}\left( \begin{array}{c} \delta X \\ \delta Y \\ \end{array}\right)
= M \left( \begin{array}{c} \delta X \\ \delta Y \\ \end{array}\right) \ ,\quad
M=\left( \begin{array}{cc} \tilde A & \tilde B \\ \tilde C & \tilde D \\
\end{array}\right)
\ee
Here
\bea
\label{GB45b}
\tilde A&\equiv &
\frac{3\gamma^2 \phi_0 X_0^2 + 48\gamma X_0^2 Z_0 - 6\gamma\phi_0
  - 208 \gamma \phi_0 X_0 Z_0 + 768\left(\phi_0 - X_0\right)Z_0^2}
{2\phi_0\left(\gamma + 8\gamma X_0Z_0 + 96 Z_0^2\right)}\ ,\nn
\tilde B&\equiv & \frac{16\gamma X_0^3 - 104\gamma \phi_0 X_0^2 + 48\phi_0
+ 1536\phi_0 X_0 Z_0 - 768X_0^2 Z_0}
{2\phi_0\left(\gamma + 8\gamma X_0Z_0 + 96 Z_0^2\right)}\ ,\nn
\tilde C&\equiv & \frac{2Z_0\left( - \gamma^2\phi_0 X_0 + 96 Z_0^2
+ \gamma + 16\gamma \phi_0 Z_0\right)}
{\phi_0\left(\gamma + 8\gamma X_0Z_0 + 96 Z_0^2\right)}\ ,\nn
\tilde D&\equiv & \frac{32 Z_0\left( -12\left(\phi_0 - X_0\right)Z_0 + \gamma 
\phi_0 X_0\right)}
{\phi_0\left(\gamma + 8\gamma X_0Z_0 + 96 Z_0^2\right)}\ \ .
\eea
If the real parts of all the eigenvalues of the matrix $M$ are negative, the
perturbation becomes
small and the system is stable. Then the condition of the stability is given by
\be
\label{GB46}
\tilde A+ \tilde D<0\ ,\quad \tilde A \tilde D - \tilde B \tilde C>0\ .
\ee
First, the check of the stability for $V=0$ case is in order.
By using (\ref{GB41}), (\ref{GB42}), and (\ref{GB43}), we find
\be
\label{GB49}
X_0^2=\frac{\phi_0^2}{h_0^2}= -\frac{6\left(h_0 - 1\right)}{\gamma
\left(5h_0 - 1\right)}\ ,\quad
Z_0^2=- \frac{\gamma \left(3h_0 - 1\right)^2}{96\left(h_0 -
1\right)\left(5h_0 - 1\right)}\ ,\quad
X_0Z_0 = - \frac{3h_0 - 1}{4 \left(5h_0 - 1\right)}\ .
\ee
In order that $X_0^2$ and $Z_0^2$ are positive, it follows
\bea
\label{GB50}
& \frac{1}{5}<h_0<1 \ , \quad
& \mbox{when}\ \gamma>0 \nn
\mbox{or}\ & h_0<\frac{1}{5}\ \mbox{or}\ h_0>1\ ,\quad & \mbox{when}\ \gamma<0\ 
.
\eea
By using (\ref{GB49}),  $\tilde A$, $\tilde B$, $\tilde C$, and $\tilde D$ in
(\ref{GB45b}) can be expressed in terms of $h_0$:
\bea
\label{GB51}
&& \tilde A = \frac{\left(h_0 - 1\right)\left(9h_0^2 - 4 h_0 + 1 \right)}
{h_0\left(5h_0^2 - 4h_0 + 1\right)}\ ,\quad
\tilde B = \frac{24\left(h_0 -1\right)\left(3h_0^2 - 2h_0 +1\right)}
{\gamma h_0 \left(5h_0^2 - 4h_0 + 1\right)}\ ,\nn
&& \tilde C = \frac{\gamma \left(3h_0 -1\right) \left(3h_0^2 -1\right)}
{12 \left(h_0 -1\right)\left(5h_0^2 - 4h_0^2 + 1\right)}\ ,\quad
\tilde D = -\frac{2\left(h_0-1\right)^2 \left(3h_0 -1\right)}
{h_0\left(5h_0^2 - 4h_0 + 1\right)}\ .
\eea
Then we obtain very simple results:
\bea
\label{GB52}
\tilde A + \tilde D &=& - \frac{3\left(h_0 - 1\right)}{h_0} \ ,\\
\label{GB53}
\tilde A \tilde D - \tilde B \tilde C &=&
\frac{2\left(1 - 3h_0\right)}{h_0^2}\ .
\eea
Therefore Eq.(\ref{GB46}) is satisfied and the system is stable if
and only if
\be
\label{GB58}
h_0<0\ .
\ee
Then the case corresponding to phantom cosmology with $h_0<0$ is always stable.

\end{document}